\documentclass[sigconf]{acmart}

\usepackage{booktabs}
\usepackage{graphicx}
\usepackage{enumitem}
\newcommand{\etal}{\textit{et al.}}
\usepackage{lipsum}
\usepackage{float}

\AtBeginDocument{%
  \providecommand\BibTeX{{%
    \normalfont B\kern-0.5em{\scshape i\kern-0.25em b}\kern-0.8em\TeX}}}

\copyrightyear{2022}
\acmYear{2022}
\setcopyright{acmlicensed}\acmConference[SIGIR '22]{Proceedings of the 45th International ACM SIGIR Conference on Research and Development in Information Retrieval}{July 11--15, 2022}{Madrid, Spain}
\acmBooktitle{Proceedings of the 45th International ACM SIGIR Conference on Research and Development in Information Retrieval (SIGIR '22), July 11--15, 2022, Madrid, Spain}
\acmPrice{15.00}
\acmDOI{10.1145/3477495.3531818}
\acmISBN{978-1-4503-8732-3/22/07}

\begin{document}
\fancyhead{}

\title{Analysing~the~Robustness~of~Dual~Encoders for~Dense~Retrieval~Against~Misspellings}

\author{Georgios Sidiropoulos}
\affiliation{%
  \institution{University of Amsterdam}
  \city{Amsterdam}
  \country{The Netherlands}}
\email{g.sidiropoulos@uva.nl}

\author{Evangelos Kanoulas}
\affiliation{%
  \institution{University of Amsterdam}
  \city{Amsterdam}
  \country{The Netherlands}}
\email{e.kanoulas@uva.nl}

\begin{abstract}
Dense retrieval is becoming one of the standard approaches for document and passage ranking. The dual-encoder architecture is widely adopted for scoring question-passage pairs due to its efficiency and high performance. Typically, dense retrieval models are evaluated on clean and curated datasets. However, when deployed in real-life applications, these models encounter noisy user-generated text. That said, the performance of state-of-the-art dense retrievers can substantially deteriorate when exposed to noisy text.
In this work, we study the robustness of dense retrievers against typos
in the user question. We observe a significant drop in the performance of the dual-encoder model when encountering typos and explore ways to improve its robustness by combining data augmentation with contrastive learning. Our experiments on two large-scale passage ranking and open-domain question answering datasets show that our proposed approach outperforms competing approaches. Additionally, we perform a thorough analysis on robustness. Finally, we provide insights on how different typos affect the robustness of embeddings differently and how our method alleviates the effect of some typos but not of others.
\end{abstract}



\maketitle

\section{Introduction}
With the advances in neural language modeling \cite{DBLP:conf/naacl/DevlinCLT19}, learning dense representations for text has become a vital component in many information retrieval (IR) tasks. In passage ranking and open-domain question answering, dense retrieval has become a new paradigm to retrieve relevant passages \cite{DBLP:conf/sigir/KhattabZ20, DBLP:conf/emnlp/KarpukhinOMLWEC20, DBLP:journals/tacl/LuanETC21}. In contrast to traditional term-based IR models (TF-IDF and BM25) that fail to capture beyond lexical matching, dense retrieval learns dense representations of questions and passages for semantic matching. 

A typical approach for dense retrieval involves learning a dual-encoder for embedding the questions and passages \cite{DBLP:conf/emnlp/KarpukhinOMLWEC20}. A dual-encoder model consists of two separate neural networks optimized to score relevant (i.e., positive) question-passage pairs higher than irrelevant (i.e., negative) ones. At inference time, the score of a question-passage pair is computed as the inner product of the corresponding question and passage embeddings. Due to their high efficiency, dual-encoders are popular first-stage rankers in large-scale settings (in contrast to cross-encoders where even though they can achieve higher performance, they are not indexable and therefore are used as re-rankers \cite{DBLP:conf/ecir/GaoDC21, sidiropoulos-etal-2021-combining}). The whole corpus can be encoded and indexed offline, while at inference time, high-scoring passages with respect to a question can be found using efficient maximum inner product search~\cite{DBLP:journals/tbd/JohnsonDJ21}.

So far, dense retrieval models have been evaluated on clean and curated datasets. However, these models will encounter user-generated noisy questions when deployed in real-life applications.
Questions can include typos because of users mistyping words, such as keyboard typos (additional/missing character and character substitution), phonetic typing errors due to the close pronunciation, and misspellings. How these typos affect the encoding of questions and whether dense dual-encoder retrieval models are robust to them is not studied yet.

Works on text classification have shown that deep neural language models such as BERT are not robust against typos \cite{DBLP:conf/acl/PruthiDL19, DBLP:journals/corr/abs-2003-04985}, even though they apply the \textit{WordPiece} tokenization. Ma~\etal{}~\cite{DBLP:conf/acl/MaSA21} and Zhuang~\etal{}~\cite{DBLP:conf/emnlp/ZhuangZ21} showed that typos can confuse even advanced BERT-based cross-encoders for re-ranking \cite{DBLP:conf/sigir/DaiC19, DBLP:journals/corr/abs-1901-04085, DBLP:conf/ecir/GaoDC21} and proposed data augmentation training for building typo-robust re-rankers. Additionally, Ma~\etal{}~\cite{DBLP:conf/acl/MaSA21} showed that bringing closer in the latent space the representations of the positive question-passage pairs of different questions while being far apart from negative ones can increase robustness.

While the aforementioned works studied robustness for the case of re-ranking, improving the robustness of dense retrieval for first-stage ranking has not been explored in-depth yet. Intuitively, if typos cause an inferior first-stage ranking, that will already negatively affect the performance of the re-ranker. Therefore, robustness for first-stage ranking is crucial for the overall performance. To the best of our knowledge, Zhuang~\etal{}~\cite{DBLP:conf/emnlp/ZhuangZ21} is the only work that studied the first-stage ranking and used data augmentation to improve the robustness of a BERT-based Siamese encoder. 

In this paper, we study in-depth the robustness of dense retrieval for the case of dual-encoder architecture. We propose an approach that combines data augmentation with contrastive learning for robustifying dual-encoders against questions with typos. In detail, alongside augmenting questions with typos, we propose to use a contrastive loss that brings the representation of a question close to its typoed variations in the latent space while keeping it distant from other distinct questions.

We aim to answer the following research questions:
\begin{enumerate}[label=\textbf{RQ\arabic*},leftmargin=*]
\item Can data augmentation, contrastive learning, and their combination improve the robustness of dense retrieval to typos?
\item Do certain typoed words affect the robustness of the question encoding more than others?
\item Do the proposed method improve the robustness of the question encoding by ways other than simply learning to ignore the typoed word?
\end{enumerate}

Our main contributions are the following: (i) we propose an approach for robustifying dense retrievers towards typos in user questions that combines data augmentation with contrastive learning and performs better than applying each component separately, (ii) we perform a thorough analysis on the robustness of dense retrieval, and (iii) show that typos in various words influence performance differently. \footnote{\url{https://github.com/GSidiropoulos/dense-retrieval-against-misspellings}.}

\section{Experimental Setup}
In this section, we discuss the datasets, the metrics, and the robustness methods we experiment with to answer our research questions.
\subsection{Datasets}
We conduct our experiments on two large-scale datasets, namely, MS MARCO passage ranking \cite{DBLP:conf/nips/NguyenRSGTMD16}, and Natural Questions \cite{DBLP:journals/tacl/KwiatkowskiPRCP19}.
In MS MARCO passage ranking, the goal is to rank passages based on their relevance to a question (i.e., the probability of including the answer). The data collection consists of 8.8 million passages; the questions were selected from Bing search logs. Natural Questions is a large-scale dataset for open-domain QA over Wikipedia, and its questions were selected from Google search logs. Table \ref{tab:datasets} shows the statistics of the two datasets.

\begin{table}[ht!]
\caption{Number of questions in each dataset, and the average length of question.}
\resizebox{\columnwidth}{!}{%
\begin{tabular}{@{}ccccc@{}}
\toprule
 & Train & Dev & Test & Avg. q length \\ \midrule
MS MARCO & 502,939 & 6,980 & 6,837 & 5.94 \\
Natural Questions & 79,168 & 8,757 & 3,610 & 9.20 \\ \bottomrule
\end{tabular}%
}
\label{tab:datasets}
\end{table}
\subsection{Metrics}
To measure the retrieval performance on MS MARCO, we use the official metric MRR (@10) alongside the commonly reported Recall (R) at top-k ranks \cite{DBLP:conf/sigir/KhattabZ20,DBLP:conf/naacl/QuDLLRZDWW21}.\footnote{Similar to previous works, we report the metrics on MSMARCO (Dev) since the correct answers for the test set are not available to the public.} Following previous work on Natural Questions, we use answer recall (AR) at the top-K retrieved passages \cite{DBLP:conf/emnlp/KarpukhinOMLWEC20, DBLP:conf/naacl/QuDLLRZDWW21}. Answer recall measures whether at least one of the top-k retrieved passages contains the ground-truth answer.

\subsection{Methods}
\label{sec:methods}
In this section, we describe the dual-encoder model we use for our experiments. Moreover, we present the three approaches we apply as extensions to this model in order to increase robustness.

\textbf{Dense Retriever (DR)} is a dual-encoder BERT-based model used for scoring question-passage pairs \cite{DBLP:conf/emnlp/KarpukhinOMLWEC20}. Given a question $q$, a relevant passage $p^+$ and a set of irrelevant passages $\{p_1^-, p_2^-, \dots, p_n^-\}$,  the model learns to rank $p^+$ higher than the negative passages via the optimization of the negative log-likelihood of the relevant passage:
\begin{eqnarray}
&& \mathcal{L}_{1}(q_i, p^+_i, p^-_{i,1}, \cdots, p^-_{i,n}) \label{eq:training} \\
&=& -\log \frac{ e^{\mathrm{sim}(q_i, p_i^+)} }{e^{\mathrm{sim}(q_i, p_i^+)} + \sum_{j=1}^n{e^{\mathrm{sim}(q_i, p^-_{i,j})}}}. \nonumber
\end{eqnarray}

\textbf{Data Augmentation (DR + Data augm.)} is one of the traditional approaches for robustifying neural models. By exposing DR on questions with and without typos, the model learns to be invariant to typos. Similar to Zhuang~\etal{}~\cite{DBLP:conf/emnlp/ZhuangZ21}, for each original correctly written question, on training time, we draw an unbiased coin. If the result is heads, we use the original question for training. If the result is tails, we use one of its typoed variations.

\textbf{Contrastive learning (DR + CL)} of representations works by maximizing the agreement between differently augmented views of the same object. We propose a contrastive loss that compares the similarity between a question and its typoed variations and other distinct questions. In contrast with data augmentation, which explicitly trains on typoed question-passage pairs, here such pairs are seen implicitly only. In detail, in addition to Equation \ref{eq:training}, we introduce a loss that enforces that a question, $q$ and its typoed variations $q^+$ are close together in the latent space while being far apart from other distinct questions $\{q_1^-, q_2^-, \dots, q_n^-\}$:

\begin{eqnarray}
&& \mathcal{L}_{2}(q_i, q^+_i, q^-_{i,1}, \cdots, q^-_{i,n}) \label{eq:contrastive} \\
&=& -\log \frac{ e^{\mathrm{sim}(q_i, q_i^+)} }{e^{\mathrm{sim}(q_i, q_i^+)} + \sum_{j=1}^n{e^{\mathrm{sim}(q_i, q^-_{i,j})}}}. \nonumber
\end{eqnarray}
The final loss is a weighted average of the two losses:
\begin{eqnarray}
    \mathcal{L} = w_{1} \cdot \mathcal{L}_{1} + w_{2} \cdot \mathcal{L}_{2}
\label{eq:overll}
\end{eqnarray}

The weights $w_1$ and $w_2$ are hyper-parameters and therefore need to be defined. Giving equal weights to the two losses is an effective and straightforward combination method which we used in our experiments.

\textbf{Combination (DR + Data augm. + CL)} method consists of data augmentation combined with contrastive learning. Specifically, we propose alongside augmenting questions with typos to use the contrastive loss of Equation \ref{eq:contrastive} that brings the representation of a question close to its typoed variations while keeping it distant from other distinct questions. The final loss is a weighted average of the three losses:
\begin{eqnarray}
    \mathcal{L} = w_{1} \cdot \mathcal{L}_{1} + w_{2} \cdot \mathcal{L}_{2}+ w_{3} \cdot \mathcal{L}_{3},
\label{eq:overll}
\end{eqnarray}
where $\mathcal{L}_{3}$ represents the data augmentation and is computed similarly to Equation \ref{eq:training}  but for the typoed variation $q^+$ of the original question $q$. For our experiments, we use an equal weighting setting for the weights $w_1$,$w_2$, and $w_3$.

\subsection{Simulating Typos}
\label{sec:simulate_typos}
To assess the robustness of the proposed methods, a large-scale dataset for passage retrieval with typoed questions is necessary. Unfortunately such a dataset does not exist, and therefore we build one by simulating typos over the original Natural Questions and MS MARCO datasets. In detail, we simulate typos produced by humans by augmenting the original questions in the dataset with synthetically generated typoed ones.

In order to simulate typos, we apply the following transformations that often occur in human-generated questions.

\begin{itemize}
\item Random: Inserts, deletes, swaps, or substitutes a random character; e.g., \textit{committee} $\rightarrow \{$\textit{copmmittee}, \textit{commttee}, \textit{comimttee}, \textit{commitlee}$\}$. 
\item Keyboard: Swaps a random character with those close to each other on the QWERTY keyboard; e.g., \textit{committee} $\rightarrow$ \textit{comnittee}.
\item Common misspellings: Replaces words with misspelled ones, defined in a dictionary of common user-generated misspellings; e.g., \textit{committee} $\rightarrow$ \textit{comittee}.
\end{itemize}

\subsection{Implementation Details}
\label{sec:details}
The DR model used in our experiments is trained using the in-batch negative setting described in \cite{DBLP:conf/emnlp/KarpukhinOMLWEC20}. The question and passage BERT encoders are trained for $50K$ steps, with a batch size of $48$. The learning rate is set to $2e$-$5$ using Adam, and the rate of the linear scheduling with a warm-up is set to $0.1$. Moreover, we use the same hyper-parameters for the three robustifying methods described in Section \ref{sec:methods}, in order to ensure a fair comparison. 

For generating typos in the training phase as well as building the typo-robustness test set, we use the open-source Aug library \cite{bitton2021augly}. Each word in a question gets transformed with a probability of 0.2, and the transformation type (Section \ref{sec:simulate_typos}) gets chosen at random.

\begin{table*}[ht!]
\caption{Retrieval results for the settings of (i) clean questions (Original), and (ii) questions with typos (Typos in Random Words). Statistical significance difference with paired t-test $(p < 0.05)$ DR=d; DR+ Data augm.=a; DR+CL=c.}
\resizebox{\textwidth}{!}{%
\begin{tabular}{@{}l|llllll|llllll@{}}
\toprule
 & \multicolumn{6}{c|}{Natural Questions (Test)} & \multicolumn{6}{c}{MS MARCO (Dev)} \\ \midrule
 & \multicolumn{3}{c}{Original} & \multicolumn{3}{c|}{Typos in Random Words} & \multicolumn{3}{c}{Original} & \multicolumn{3}{c}{Typos in Random Words} \\
 & AR@5 & AR@20 & AR@100 & AR@5 & AR@20 & AR@100 & MRR@10 & R@50 & R@1000 & MRR@10 & R@50 & R@1000  \\ \midrule
DR & 67.31 & 78.22 & 85.42 & 49.52 & 63.98 & 76.12 & 28.11 & 73.46 & 93.36 & 15.11 & 46.47 & 74.02 \\
DR + Data augm. & 66.45 & 79.03 & 85.56 & 60.69 & 73.76 & \textbf{83.37} & 28.26 & 72.66 & 93.07 & 22.00 & 61.68 & 86.49 \\
DR + CL (ours) & 66.31 & 77.42 & 85.45 & 55.51 & 69.27 & 80.52 & 28.95 & 73.01 & 93.64 & 19.37 & 55.08 & 80.69 \\
\begin{tabular}[c]{@{}l@{}}DR + Data augm.\\ + CL (ours)\end{tabular} & 67.47 & 78.83 & 85.67 & \textbf{62.13}$^{dac}$ & \textbf{74.87}$^{dac}$ & 83.26$^{dc}$ & 29.14 & 73.85 & 93.69 & \textbf{22.84}$^{dac}$ & \textbf{63.21}$^{dac}$ & \textbf{87.52}$^{dac}$ \\ \bottomrule
\end{tabular}%
}
\label{tab:rq1}
\end{table*}

\begin{table*}[ht!]
\caption{Retrieval results for the settings of (i) questions with typos in non-stopwords (Typos in Non-stopwords), and (ii) questions with typos in highly discriminative utterances (Typos in Discriminative Utterances). Stat. sig. difference w/ paired t-test $(p < 0.05)$ DR=d; DR+Data augm.=a; DR+CL=c.}
\resizebox{\textwidth}{!}{%
\begin{tabular}{@{}l|llllll|llllll@{}}
\toprule
 & \multicolumn{6}{c|}{Natural Questions (Test)} & \multicolumn{6}{c}{MS MARCO (Dev)} \\ \midrule
 & \multicolumn{3}{c}{Typos in Non-stopwords} & \multicolumn{3}{c|}{Typos in Discriminative Utterances} & \multicolumn{3}{c}{Typos in Non-stopwords} & \multicolumn{3}{c}{Typos in Discriminative Utterances} \\
 & AR@5 & AR@20 & AR@100 & AR@5 & AR@20 & AR@100 & MRR@10 & R@50 & R@1000 & MRR@10 & R@50 & R@1000 \\ \midrule
DR & 40.60 & 55.48 & 68.72 & 38.89 & 53.37 & 68.00 & 11.83 & 38.98 & 66.16 & 10.51 & 34.17 & 59.71 \\
DR + Data augm. & 56.14 & 69.94 & 80.19 & 51.68 & 66.12 & 78.08 & 18.51 & 54.82 & 81.92 & 16.51 & 49.06 & 77.47 \\
DR + CL (ours)& 49.47 & 64.12 & 76.48 & 44.04 & 59.00 & 72.43 & 15.44 & 46.69 & 73.27 & 12.44 & 39.17 & 66.69 \\
\begin{tabular}[c]{@{}l@{}}DR + Data augm.\\ + CL (ours)\end{tabular} & \textbf{57.78}$^{dac}$ & \textbf{70.77}$^{dac}$ & \textbf{81.38}$^{dac}$ & \textbf{53.15}$^{dac}$ & \textbf{67.28}$^{dac}$ & \textbf{78.61}$^{dac}$ & \textbf{19.47}$^{dac}$ & \textbf{56.22}$^{dac}$ & \textbf{83.61}$^{dac}$ &  \textbf{17.58}$^{dac}$ & \textbf{50.81}$^{dac}$ & \textbf{79.51}$^{dac}$ \\ \bottomrule
\end{tabular}%
}
\label{tab:rq2}
\end{table*}

\section{Results}
In this section, we present our experimental results that answer our research questions. We aim to answer \textbf{RQ1} by comparing the retrieval performance of the methods we consider (Section \ref{sec:methods}) for two settings: clean questions and questions with typos. In Table \ref{tab:rq1}, we obverse that on clean questions, data augmentation as well as our two proposed approaches, namely, contrastive learning and data augmentation combined with contrastive learning do not harm the performance. Moreover, all the approaches for robustifying DR are performing significantly better compared to the original DR, on questions with typos. That indicates that they successfully robustify the underlying dual-encoder model. That said, our proposed data augmentation combined with contrastive learning approach holds the best performance.

Following previous works, we randomly introduce typos to questions. However, we want to investigate if the performance of the approaches we consider remains the same irrespectively of the word in which the typo appears. To answer \textbf{RQ2}, we create two additional test settings for the case of questions with typos. Specifically, we create (i) a setting where typos appear only in non-stopwords, and (ii) a setting where typos appear only in utterances with lexical match with the relevant passage.\footnote{We build the new settings using the same probability for introducing typos (Section \ref{sec:details}), and we do not retrain the models on the new settings.} In detail, we consider the overlapping consecutive words between the ground-truth passage and the question (e.g., ``Who was the president of the united states during wwi?'', and ``Woodrow Wilson, a leader of the Progressive Movement, was the 28th President of the United States (1913-1921). After a policy of neutrality at the outbreak of World War I, he led America into war.'' mark the ``president of the united states'' as available utterance to introduce typos). The highly discriminative utterances obtained through this heuristic are typically entity mentions.

As we can see in Table \ref{tab:rq2} and by comparing the numbers with the results in Table \ref{tab:rq1}, the effectiveness of the methods varies across the three settings. Particularly, robustness deteriorates when typos do not appear randomly. In detail, the most significant losses occur when typos appear on discriminative utterances. To this extent, our proposed data augmentation combined with contrastive learning approach remains the best performing one across all setting.

\begin{figure}[H]
        \centering
        \includegraphics[width=1\linewidth]{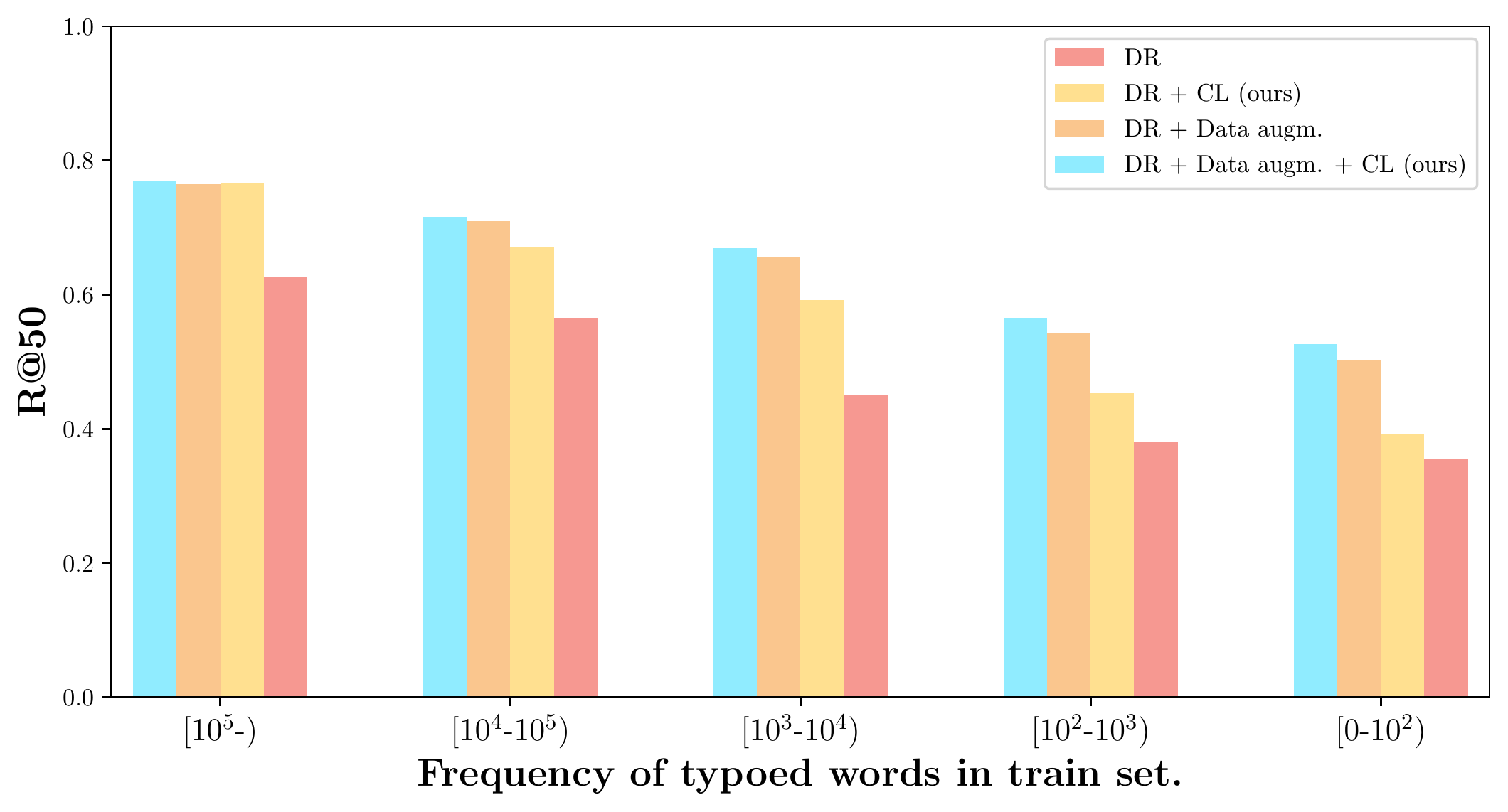}
        \caption{Retrieval performance (Average Recall@50) w.r.t the frequency on the training set, of the typoed words at test-time; on MS MARCO (Dev). Questions are split into bins w.r.t the frequency of their typoed words.}
        \label{fig:tf_typos}
\end{figure}

To better understand the discrepancy in robustness between the three settings of questions with typos, we conduct the following analysis. For the setting where typos randomly appear on questions, we study how the frequency on the training set of the typoed words at test time affects robustness. As shown in Figure \ref{fig:tf_typos}, there is a strong connection between the frequency of the typoed words and the retrieval performance. As the frequency of the typoed words decreases, the performance drops significantly. To this extent, our proposed data augmentation combined with contrastive learning approach remains the best performing one, with the performance gap increasing as the frequency of the typoed word decreases. The results in Figure \ref{fig:tf_typos} can also explain why we observe the highest losses in performance on the setting with typos in discriminative utterances. In general, the discriminative utterances (entity mentions) diversity between the dataset splits is higher compared to other words appearing in questions (e.g., interrogative, linking words).

For \textbf{RQ3}, we consider a simple baseline where the typoed words are identified and removed from the question before being fed to the original DR model. We compare our best-performing approach against the aforementioned baseline. If our proposed approach only learns to ignore words with typos, then we argue that the performance of the baseline should be competitive to ours. In many cases, removing the typoed word can be a valid approach since the importance of words in a question varies. For instance, considering the question ``Where was president Lincoln born?'' we see that ``Lincoln'' is crucial for the meaning of the question while ``was'' adds no information. With that in mind, we study \textbf{RQ3} with respect to the relative importance of the typoed words within the question.\footnote{We define a word's relevant importance as the ratio of its IDF to the sum of the IDFs of every word in the question.}

\begin{figure}[H]
        \centering
        \includegraphics[width=1\linewidth]{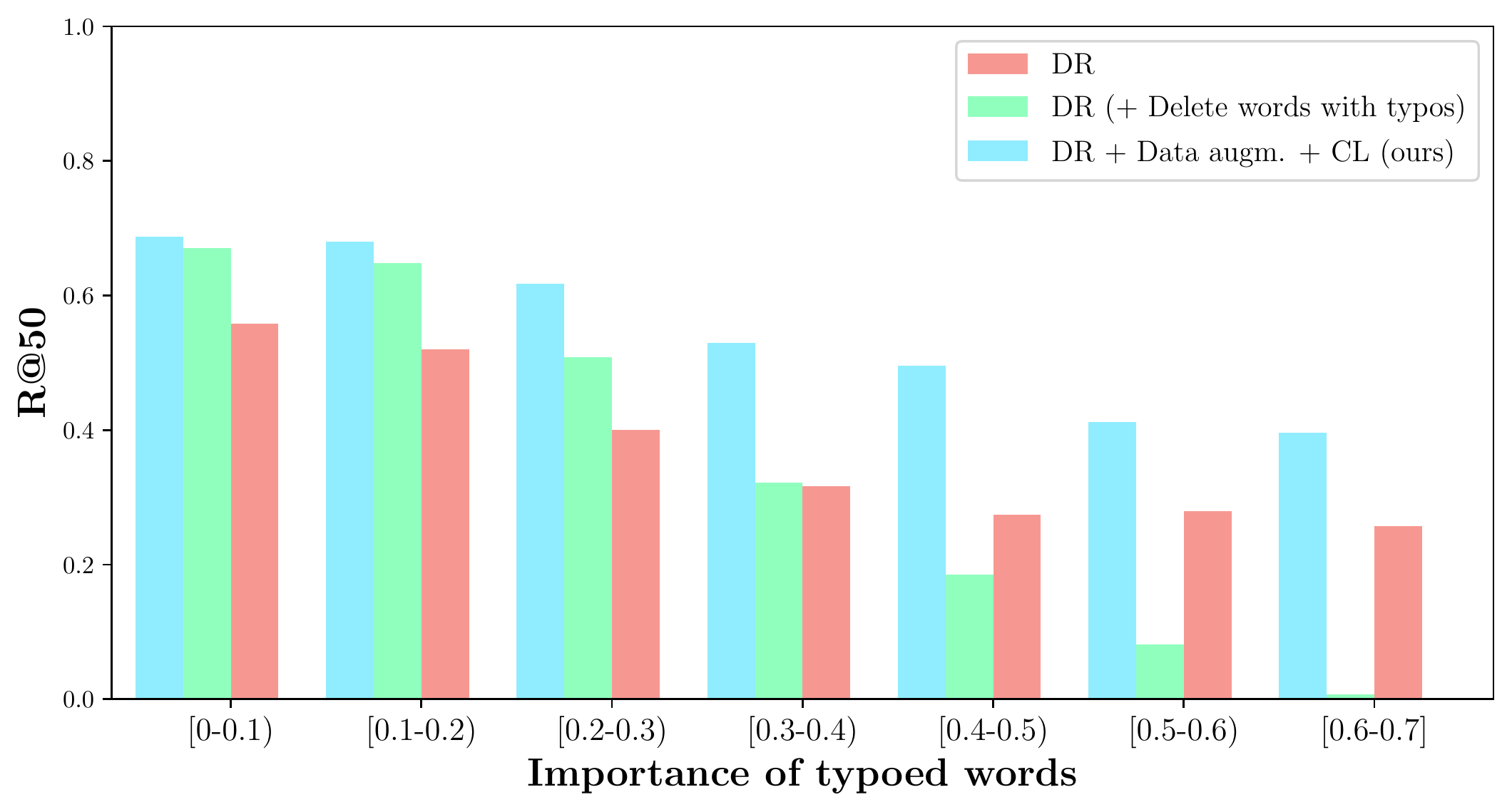}
        \caption{Retrieval results w.r.t the relevant importance of the typoed words; on MS MARCO (Dev). Questions are split into bins w.r.t the relevant importance their typoed words.}
        \label{fig:idf_typos_remove_baseline}
\end{figure}

Our method does not simply learn to ignore words with typos since, as we can see from Figure \ref{fig:idf_typos_remove_baseline}, it consistently outperforms the baseline. Furthermore, Figure \ref{fig:idf_typos_remove_baseline} highlights that when the importance of typoed words is low, simply ignoring them is a highly competitive approach. On the other hand, as the importance of the typoed words increases, the effectiveness of just ignoring these words decreases dramatically, to the extent that keeping the typoed words performs better. That can be attributed to the application of the WordPiece tokenizer (by the underlying BERT model) that allows DR to recover from some typos, such as when the character n-gram splits remain intact despite the typos. For instance original \textit{robustness} and typoed \textit{robustnessd} will be split into \textit{[robust, $\#\#$ness]} and \textit{[robust, $\#\#$ness, $\#\#$d]} respectively.

\section{Conclusions}
In this work, we provided insights on the robustness of dual-encoders to typos of user questions for dense retrieval. We proposed an approach for robustifying dual-encoders that combines data augmentation with contrastive learning. Our experimental results showed that our proposed method not only improves robustness but also performs better than separately applying data augmentation or contrastive learning. Analysis of the methods we explored showed that typos in various words do not influence performance equally. In particular, typos on words that are less frequent on the training set and more important for a question are harder to address. Our proposed technique remains the best performing one in these settings, however the performance deteriorates significantly compared to a clean question. There is a significant question that has risen throughout our study: What could a dual-encoder actually learn to fix the problem? The \textit{WordPiece} tokenizer, applied by BERT, allows models to recover from some typos. However, it would be ideal if embeddings could be learned at the character n-gram level to allow recovery from typical character substitution, deletion, etc. Furthermore, word-to-word interactions during training (e.g., through a late interaction model~\cite{DBLP:conf/sigir/KhattabZ20}) could also allow implicitly to learn the ``correct spelling'' of a typoed word during training. We leave these directions as future work.

\begin{acks}
This research was supported by
the NWO Innovational Research Incentives Scheme Vidi (016.Vidi.189.039),
the NWO Smart Culture - Big Data / Digital Humanities (314-99-301),
the H2020-EU.3.4. - SOCIETAL CHALLENGES - Smart, Green And Integrated Transport (814961).
All content represents the opinion of the authors, which is not necessarily shared or endorsed by their respective employers and/or sponsors.
\end{acks}

\bibliographystyle{ACM-Reference-Format}
\bibliography{main}

\end{document}